\documentclass{PoS}
\usepackage{natbib}

\title{What PhD students really want}

\ShortTitle{What PhD students really want}

\author{\speaker{Minnie Mao}%
\\
        University of Tasmania, ATNF, AAO\\
        E-mail: \email{minnie.mao@csiro.au}}

\abstract{The road to becoming an astronomer is exciting, but often fraught with danger and conflicting messages. A PhD student is inundated with catch-phrases such as ``publish or perish'' and ``it's not about the quantity, but the quality of work''. How do we know which advice to follow? How can we publish copious amounts of quality work in only three years so as to maximize our success in the future? How do we even know what ``good quality'' really is? With only a short time to prepare ourselves for the big wide world of Astronomy, what is the best way for a PhD student to maximize their research and ultimately maximize their success as a real astronomer?

The PhD students of today are the astronomers of tomorrow, but their journey depends on  a positive work environment in which they can thrive and improve. Here I present the results of a survey of current PhD students on how they believe they can maximize their success in science.  I  find that PhD students in Australia  expect to write more papers during their PhD than is expected by their supervisors, but that they are generally happy with the quality of their supervision. Above all, students love  telescopes, and hands-on observations are an important part of acquiring the knowledge and culture necessary to becoming a real astronomer.}

\FullConference{Accelerating the Rate of Astronomical Discovery\\
                 August 11-14 2009\\
                 Rio de Janeiro, Brazil}

\begin{document}

\section{Introduction}
In Australia, PhD students in astronomy are funded for three years with a possible extension of six months. In other words, astronomy PhD students in Australia have three years to become world experts in their field of research and make themselves desirable employees. As anyone who has survived a PhD knows, there is a lot of well-meaning advice to help PhD students along this danger-fraught path. The perils of not publishing enough papers is constantly reiterated through the alliterative ``publish or perish''. However, the pitfalls of not producing a quality thesis are also stated over and over again, with good reason. It is therefore imperative that students manage their short PhD-span efficiently so that they can publish the requisite number of papers, write a good thesis and also balance this with a healthy social life. Moreover, it is vital that they enjoy their time as PhD students. Not only are happy people more productive \citep[e.g.][]{Seligman98}, enthusiastic students will grow into passionate supervisors who will in turn nurture the next generation of enthusiastic students!

How many papers should a student publish during their PhD? What motivates them? What do the students think is most important for their PhDs? This paper presents the results of a survey I conducted of 39 astronomy PhD students in Australia on what they thought was most important to their PhD. I also surveyed seven staff astronomers. This work was conducted in a Universe where we assume the standard $\Lambda$CDM cosmology. 

\section{Australian PhDs}
This paper is based on Australian PhD students. A PhD in astronomy in Australia is typically funded for 3 - 3.5 years and is generally 100\% research. Some universities have a coursework component but it is not assessed and is not a requirement for a PhD. The Australian system differs from some other countries in that between the undergraduate Bachelors degree and the PhD there is a year-long program called ``Honours''\footnote{Some universities offer a ``Masters'' program as an alternative to Honours.}. Honours comprises coursework and research. PhD candidates will only be accepted if they do sufficiently well in Honours. 

\section{The survey}
The survey participants' details are shown in Table \ref{surveysumm} while the survey questions and results are shown in Table \ref{survey}. We emphasize that the survey included a lot of space for comments so participants were free to discuss any other issues that they felt were important. Some of these are discussed in Section \ref{unexpected}. 


\begin{table}
\begin{center}
\caption{Details of the survey participants.  One of the students did not specify their gender. Although many of the survey participants did both observational and theoretical work, for the purposes of this survey only theorists with no observing component are shown as ``theorists''.}
\begin{tabular}{ccc}
\\
\hline
& Students & Staff\\
\hline
Female & 18 & 1\\
Male & 20 & 6\\
First year PhD& 17 & -\\
Second year PhD& 14 & -\\
Third year PhD& 6 & -\\
Fourth year PhD& 2 & -\\
Observers & 35 & 7\\
Theorists & 4 & 0\\
\hline
\end{tabular}\label{surveysumm}
\end{center}
\end{table}

The survey was conducted on students and speakers at the Harley Wood Winter School (HWWS) of 2009\footnote{http://asa2009.science.unimelb.edu.au/HWWS/}, which is an annual three day workshop, sponsored by the Astronomical Society of Australia\footnote{http://asa.astronomy.org.au/}, held for astronomy students. The nature of the school prevents the surveyed students and staff from representing a complete sample. For example, the surveyed students are more likely to be local, newer, and keener.   A total of 39 PhD students (``students''), and 7 staff astronomers (``staff'')  completed the survey.



\subsection{Gender}
It is interesting to speculate that the gender imbalance of staff may be due to the gender imbalance said to be prevalent in previous generations but I remain wary of small number statistics. It is however gratifying to note the gender balance of students.

\subsection{Stage-of-PhD}
Out of the 39 surveyed PhD students, 31 ($\sim$86\%) are in the first or second year of their PhD. This is most likely due to the HWWS being targeted towards newer students. Furthermore, the PhD students in their third and fourth years are probably desperately trying to write up their theses and do not have time to complete surveys.

\subsection{Observer or Theorist?}
The majority of the students identified themselves as observers with only four ($\sim$10\%) identifying themselves as pure theorists. All the staff  were observers. The survey allowed participants to identify themselves as observers, theorists or both. Although only four participants were solely theorists, many students considered themselves to be observers and theorists.

\section{Results \& Discussion}\label{discussion}

\begin{table}
\begin{footnotesize}
\begin{center}
\caption{The survey questions and responses. For the first ten statements we present the average response with the standard deviation as the error. For Questions 1 - 4, a response of 1 corresponds to  ``strongly disagree'' and 5 corresponds to ``strongly agree''. For Questions 5 - 10, a response of 1 corresponds to ``not at all important'' and 5 corresponds to ``extremely important''. Questions 11 and 12 were yes/no questions. Question 13 again presents the average response with the standard deviation as the error. }
\begin{tabular}{llcc}

\\
\hline
&Statement/Question & Students & Staff\\
\hline
1&My supervisor/s spends too little time with me & 2.3  $\pm$ 1.1\\
2&My supervisor/s restrict my research freedom & 1.7 $\pm$ 0.8\\
3&My supervisor/s provide me with enough direction & 4.0 $\pm$ 0.8\\
4&My supervisor/s motivate and inspire me & 4.1 $\pm$ 1.0\\
\\
5&Meetings with supervisor/s & 4.2 $\pm$ 0.9 & 4.3 $\pm$ 0.5\\
6&Summer/Winter schools, interacting with other PhD students etc & 3.9 $\pm$ 1.0 & 3.1 $\pm$ 1.3\\
7&Observing and interacting with observatory staff and other observers & 3.8 $\pm$ 1.3 & 4.3 $\pm$ 1.0\\
8&The experience of writing journal papers & 4.2 $\pm$ 1.0 & 3.8 $\pm$ 0.9\\
9&The career value of having journal papers & 4.0 $\pm$ 1.2 & 3.7 $\pm$ 1.0\\
10&International Conferences & 4.0 $\pm$ 0.8 & 3.8 $\pm$ 0.7\\
\\
11&Do you think Coursework ought to be part of a PhD? & Y: 19 N: 20 & Y: 3 N: 4 \\
12&Is going to observatories and/or the experience of observing important to you? & Y: 36 N: 3 & Y: 7 N: 0 \\
13& How many first author papers do you think is reasonable for a PhD student in \\
& Australia to have written by the end of their PhD? &3.1 $\pm$ 0.5 & 2.2 $\pm$ 0.6\\

\hline

\end{tabular}\label{survey}
\end{center}
\end{footnotesize}
\end{table}

The survey questions were designed to determine what Australian PhD students in astronomy think of their supervision; what PhD students think is important to their PhDs; what PhD students think of coursework; how important on-site observing is for PhD students, and how many first-author refereed papers PhD students are aiming to write in the duration of their PhD. The survey results are shown in Table \ref{survey}. 


\subsection{Supervision}
The first four questions addressed quality of supervision. In general I find that Australian PhD students in astronomy are happy with their level of supervision, although some students felt they were not spending enough time with their supervisors.



\subsection{Coursework}

It was surprising to find that $\sim$50\% of the students were in favour of coursework. A common theme in the comments was that three years is too short a period to complete the requisite research for a PhD and complete coursework. 

\subsection{Observing}

The response to observing was overwhelming. Every single observer, and one of the theorists, expressed that observing and visiting observatories, was important to them. Two of the three students that answered `No' left comments explaining that, as theory students, observing was ``not relevant''. Most of the survey participants responded to this question with comments passionately expounding the merits of observing. Observing was described as being ``critical'' as it gives you a ``better understanding of data collection limitations''. The observing experience was described as ``one of the best bits'' of astronomy. One of the students stated, ``If you are going to be in the field of astronomy, (even if it is theoretical), I believe it is essential to gain experience in observing in order to gain an understanding of the tools used in astronomy for gathering data.'' 

While it is wonderful to see how passionate students are about observing at telescopes, it is also worrying because hands-on observing is not feasible for the next generation of telescopes (e.g. GMT, SKA, ALMA). As many of the students' responses stated, understanding your telescope is vital to producing good science. Without the hands-on observing, how will students learn to understand their telescope? 


To ensure students gain an understanding of their telescope and hence their science, perhaps smaller observatories may be used by Universities to train their students? For example, the University of Tasmania runs three radio telescopes, the 26m Mt Pleasant Radio Telescope, the 14m Telescope and the 30m Ceduna Radio Telescope. Astronomy students are often called upon to help with observations and are encouraged to conduct their own work using these telescopes \citep[e.g.][]{Breen07}. This hands-on experience gives University of Tasmania students a good appreciation of the end-to-end operation of radio telescopes, providing an excellent preparation for observing at different observatories, such as the ATCA (Phil Edwards, Officer-In-Charge of ATCA, 2007, priv. comm.).

Hands-on observing is important on many levels. It teaches astronomers about their telescope and hence gives them a deeper understanding of their data. Furthermore, the astronomers \emph{enjoy} observing. It is one of the ``perks'' of the job. It is fun, inspirational and motivational.  It is imperative that hands-on observing is offered in some way, shape or form so as to enable the next generation of innovative astronomy. 

\subsection{Journal Papers}
Questions regarding journal papers were posed to the survey participants because I was interested in whether the students and staff had the same expectations in a PhD. Question 13 in Table \ref{survey} shows that students are aiming, on average, to write 3.1 $\pm$ 0.1 papers in the duration of their PhD, while staff are only expecting 2.2 $\pm$ 0.2 papers from their students, where the quoted uncertainties are standard errors, rather than the standard deviations given in the Table. This survey clearly shows that students and staff have significantly different expectations when it comes to writing journal papers. 

Does this mean that too much pressure is being put on students? Or is it that students have unrealistic expectations? Are students focusing too much on ``publish or perish'' to the detriment of quality science? 

A PhD student's role is to acquire the skills to become an astronomer. This involves not only becoming an expert in their field of research, but also learning how to communicate their results effectively. Students are constantly reminded of the dire consequences of not publishing enough in the academic world. Perhaps we should do away with the time consuming business of writing a thesis and instead focus on producing a few, quality, peer-reviewed papers instead. In many countries (e.g. The Netherlands), PhD students may submit a thesis comprising only papers they have written. The process of writing a paper teaches the student how to effectively communicate their science results.

\subsection{A quick word on other issues}\label{unexpected}
The final question on the survey asked for ``any other comments you'd like to make that aren't covered by this questionnaire''. Two recurring issues cropped up: funding, and being a woman in astronomy. Here I will only address the women in astronomy issue.

Many female students commented upon how difficult it is to get back into research if they take time off for family reasons (e.g. having babies). This is a strong deterrent for women who would otherwise like to be astronomers. To an extent this also ties into the ``publish or perish'' attitude in astronomy. Is there a way we can make astronomical research more ``family-friendly''? (This is also applicable to men). Perhaps instead of simply focusing on a total number of publications, prospective employers should take into account the time available for research, and should also use a measure such as the H-index \citep{Hirsh05} which measures both the scientific productivity and scientific impact of a scientist. After all, it's not the quantity, but the quality of science that really matters!


\section{Final Comments}

This survey was conducted in order to determine what PhD students want, so as to best maximize their success when they grow into real astronomers. So what do PhD students want? They want to do research and obtain good science results. They want to convey these results to the community as scientific papers. They want hands-on experience with telescopes, they really love observing. They also want free food \citep{Cham98}. 

\section*{Acknowledgements}

MYM would like to thank the IAU, IUPAP and UTas for providing the financial means of attending the IAU, as well as Jorge Cham who kindly allowed me to use his comics in my talk. MYM is grateful to Ray Norris, Melanie Johnston-Hollitt, Rob Sharp, Jim Lovell, Jamie Stevens, Andrew Hopkins, Dominic Schnitzeler and Susan Feteris for many lengthy discussions concerning this work. Finally, a gigantic thank you to all the lovely survey respondents, and to my fellow PhD-buddies Kate Randall and Elizabeth Mahony for their invaluable opinions and comments!

\end{document}